\documentclass[prl,twocolumn,showpacs]{revtex4}
\usepackage{graphicx}

\begin{document}

\newcommand{\ket}[1]{\mbox{$|\!#1\;\!\rangle$}}
\newcommand{\aver}[1]{\mbox{$<\!#1\!\!>$}}
\def\ua{\uparrow}
\def\da{\downarrow}

\title{Semiconductor few-electron quantum dot operated as a bipolar spin filter}

\author{R. Hanson, L.M.K. Vandersypen, L.H. Willems van Beveren, J.M. Elzerman, I.T. Vink, L.P. Kouwenhoven}
\affiliation{Kavli Institute of Nanoscience Delft and ERATO Mesoscopic
Correlation Project, PO Box 5046, 2600 GA Delft, The Netherlands}

\date{\today}

\begin{abstract}
We study the spin states of a few-electron quantum dot defined in a two-dimensional electron gas, by applying a large in-plane magnetic field. We observe the Zeeman splitting of the two-electron spin triplet states. Also, the one-electron Zeeman splitting is clearly resolved at both the zero-to-one and the one-to-two electron transition. Since the spin of the electrons transmitted through the dot is opposite at these two transitions, this device can be employed as an electrically tunable, bipolar spin filter. Calculations and measurements show that higher-order tunnel processes and spin-orbit interaction have a negligible effect on the polarization.
\end{abstract}


\maketitle

%INTRODUCTION%%%%%%%%%%%%%%%%%%%%%%%%%%%%%%%%%%%%%%%%%%%%%%%%%
The spin degree of freedom of electrons has great potential as a carrier of classical information (spintronics)  \cite{Wolf} and quantum information (spin quantum bits) \cite{LossDiVincenzo}. Spintronics requires a device that can filter electrons by their spin orientation, i.e. a spin filter. As for quantum information, spin filters can be used for initialization and read-out of spin quantum bits \cite{LossDiVincenzo,Lieven}. Moreover, they are an important ingredient of recent proposals to measure Bell's inequalities with entangled electron spins \cite{Saraga}.

Much experimental progress has been made using magnetic semiconductors as spin filters \cite{SCInj}. However, many recent proposals focus on spin filtering in a two-dimensional electron gas (2DEG) \cite{Recher,SpinCurrent}, since this allows easy integration with other devices such as electron spin entanglers \cite{Saraga}. Spin-dependent electron transport through a 2DEG with giant Zeeman splitting of the lowest 2D subband was recently measured \cite{MolenkampPRL}.
Also, quantum dots formed within 2DEGs have been shown to act as spin filters, by utilizing universal conductance fluctuations controlled by gate voltages \cite{JoshScience}, and via spin-dependent coupling to the leads in a perpendicular magnetic field \cite{Ciorga}. In the former case, the filtering efficiency (up to 70\% in Ref. \cite{JoshScience}) and polarity rely on the chaotic character of the dot. In the latter case, the formation of edge channels in the leads yields reproducible spin-selectivity with a high efficiency, but the polarization always corresponds to the ground state spin-orientation in the leads, and therefore the filter cannot be bipolar. The same is true for quantum point contacts, which have been used as unipolar spin filters \cite{JoshScience,PotokPRL}. For most purposes, however, a filter is required which is both bipolar and has a very high efficiency.

Recher \textit{et al.} \cite{Recher} proposed to employ the discrete spin-resolved energy levels of a quantum dot for spin filtering in a 2DEG. The low-bias electron transport through such a dot will be almost completely polarized if the Zeeman energy is much larger than the thermal energy. Furthermore, for simple pair-wise spin filling of orbital states, the polarization is opposite at any two transitions with successive electron number. The filter polarity can thus easily be reversed by changing gate voltages.

This spin filter has not been realized experimentally. The challenge is to demonstrate spin-splitting of orbital levels in a quantum dot for successive electron transitions, and to show that electrons transported at these transitions carry opposite spin. Direct measurement of the Zeeman splitting of the orbital states at the 0$\leftrightarrow$1 electron transition was reported by two groups recently \cite{ZeemanRel,PotokPRL2003}. In dots containing more than one electron, Zeeman energy has, up to now, only been observed indirectly, by comparing the energy shifts of the ground state induced by a magnetic field for successive electron numbers \cite{WeisSasaki}.

%THIS WORK%%%%%%%%%%%%%%%%%%%%%%%%%%%%%%%%%%%%%%%%%%%%%%%%%
In this work, we study the spin states of a one- and two-electron quantum dot directly, by applying a large magnetic field in the plane of the 2DEG. This field induces a large Zeeman splitting, but has a negligible effect on the orbitals in the dot. Due to the small size of the dot, the exact number of electrons is known and the orbital energy levels are well separated. Thus, we can unambiguously identify both the orbital and the spin part of the electron wave functions.  Our measurements clearly show Zeeman splitting of the two-electron triplet states. Furthermore, we observe the single-electron Zeeman splitting at the 0$\leftrightarrow$1 electron transition as well as at the 1$\leftrightarrow$2 transition. Since the two-electron ground state is a spin-singlet, this implies that the spin orientation of transmitted electrons is opposite at these two transitions. Thus, our measurements constitute the demonstration of the  spin filter proposed by Recher \textit{et al}. The influence of higher-order tunneling and spin-orbit interaction on the filter efficiency is discussed and estimated both from calculations and measurements.

%SAMPLE%%%%%%%%%%%%%%%%%%%%%%%%%%%%%%%%%%%%%%%%%%%%%%%%%%%%%%%%%%%%%%%
The quantum dot is defined in a GaAs/AlGaAs heterostructure, containing a 2DEG 90 nm below the surface with an electron density ${n_{s}=2.9\times 10^{11}}$ cm${^{-2}}$. The measurements are performed in a dilution refrigerator at base temperature \textit{T} = 20 mK with a magnetic field $B_{/\!/}$ applied in the plane of the 2DEG. The same device was used in previous experiments on a one-electron dot \cite{ZeemanRel}. The results presented here were reproduced with a similar device fabricated on a wafer with a 2DEG 60nm below the surface.

%N=1 %%%%%%%%%%%%%%%%%%%%%%%%%%%%%%%%%%%%%%%%%%%%%%%%
We first consider electron transport through the dot at the 0$\leftrightarrow$1 electron transition and show that current through the ground state is spin-polarized for $B_{/\!/}\!\neq$ 0. Since the energy separation of the orbital levels ($\Delta E_{orb}\!\approx\!$ 1.1 meV) in this device is larger than the largest source-drain bias, $eV_{SD}$, applied in the experiments, the one-electron orbital excited states are ignored. The orbital ground state is spin-degenerate at $B_{/\!/}\!=\!$ 0. In a finite magnetic field, the two spin states, parallel ($\ua$, or spin-up) and anti-parallel ($\da$, or spin-down) to the applied field, acquire a different Zeeman energy and the orbital ground state splits: $E_{\da}\!=\!E_{\ua}\!+\!\Delta E_{Z}$, with $\Delta E_{Z}\!=\! g \mu_B B_{/\!/}$.

\begin{figure}[t]
\includegraphics[width=3.3in]{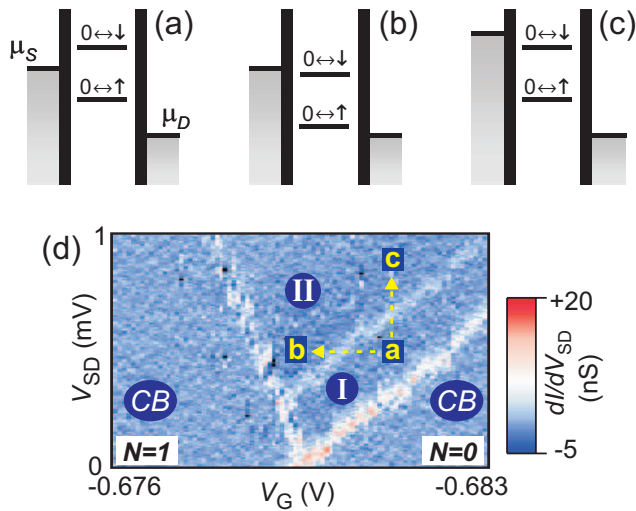}
\caption{(a)-(c) Energy diagrams showing (a) spin filtering at the 0$\leftrightarrow$1 electron transition, which can be lifted by either (b) changing the gate voltage, $V_G$, or (c) increasing the source-drain bias, $V_{SD}=(\mu_S\!-\!\mu_D)/e$. (d) $dI/dV_{SD}$ as a function of $V_G$ and $V_{SD}$ around the 0$\leftrightarrow$1 electron transition at $B_{/\!/}\!=\!$ 12 T \cite{DiamondSkewness}. In the entire region I the dot acts as a spin filter, allowing only spin-up electrons to flow through the dot. Letters a-c indicate the level positions depicted by the diagrams in (a)-(c).}
\label{fig:N=1data}
\end{figure}

Electron transport through the dot is governed by the electrochemical potential $\mu$. Consider a transition between two states of the dot, state \ket{\,a} with $N$ electrons and \ket{\,b} with $N\!\!+\!\!1$ electrons. The corresponding electrochemical potential $\mu_{a \leftrightarrow b}$ is given by the difference between the total energy of the dot in state \ket{\,a} and in state \ket{\,b}: $\mu_{a \leftrightarrow b}=U(b)-U(a)$. Choosing the zero of energy conveniently, this gives $\mu_{0\leftrightarrow \ua}\!=\!E_{\ua}$ and $\mu_{0\leftrightarrow \da}\!=\!E_{\da}\!=\!E_{\ua}+ \Delta E_{Z}$ for the 0$\leftrightarrow$1 electron transitions.

The ladder of electrochemical potentials in the dot can be shifted relative to the electrochemical potentials of the source ($\mu_S$) and the drain ($\mu_D$), by changing the gate voltage $V_G$: $\Delta \mu \propto \Delta V_G$ \cite{LeoFewEl}. Since the electrochemical potentials all depend in the same way on $V_G$, the \textit{relative} positions of the electrochemical potentials are independent of $V_G$. Thus, by tuning $V_G$, we can selectively position $\mu_{0\leftrightarrow \ua}$ in the bias window (i.e. $\mu_S\!>\!\mu_{0\leftrightarrow \ua}\!>\!\mu_D$), allowing transport of electrons through the dot via the ground state $\ket{\ua}$ only. This situation is depicted in Fig. \ref{fig:N=1data}a. Since only electrons with spin-up can enter the dot, the dot acts here as a spin filter. If the levels are pulled down by a change in $V_G$ (Fig. \ref{fig:N=1data}b), or if the source-drain bias is increased (Fig. \ref{fig:N=1data}c), transport through the spin-excited state $\ket{\da}$ becomes possible as well and the current is no longer spin-polarized. Thus, in an energy window set by the Zeeman splitting, the current through the device is carried, to first order, only by spin-up electrons, even though the leads are not spin-polarized. The influence of higher-order tunnel processes and spin-orbit interaction on the spin polarization is discussed below.

%N=1 DATA%%%%%%%%%%%%%%%%%%%%%%%%%%%%%%%%%%%%%%%%%%%%%%%%%
Fig. \ref{fig:N=1data}d shows the differential conductance $dI/dV_{SD}$ around the 0$\leftrightarrow$1 electron transition at $B_{/\!/}\!=\!$ 12 T. The lines of high $dI/dV_{SD}$ define four regions. In the regions indicated by $CB$, Coulomb blockade prohibits first-order tunneling and the number of electrons on the dot, $N$, is stable. Coulomb blockade is lifted whenever $\mu_{0\leftrightarrow \ua}$ is in the bias window, defining the V-shaped area of transport. The Zeeman splitting between $\ket{\ua}$ and $\ket{\da}$ is clearly resolved ($\Delta E_Z\approx$ 240 $\mu$eV), allowing us to identify the region where only spin-up electrons can enter the dot (region I). In region II both spin-up and spin-down electrons can pass through the dot. Thus, for all combinations of $V_G$ and $V_{SD}$ within region I, the dot acts as a spin filter transmitting, to first order, only spin-up electrons.

%***********N=2*******************
Now we analyze the current at the 1$\leftrightarrow$2 electron transition and show that in this case the dot filters the \textit{opposite} spin. The ground state of a two-electron dot in zero magnetic field is always a spin-singlet (total spin quantum number $S\!=\!$ 0) \cite{Ashcroft}, formed by the two electrons occupying the lowest orbital with their spins anti-parallel: $\ket{\:S}\!=\!(\ket{\ua\da}\!-\!\ket{\da\ua})/\sqrt{2}$. The first excited states are the spin-triplets ($S\!=\!$ 1), where the antisymmetry of the two-electron wave function requires one electron to occupy a higher orbital. The three triplet states are degenerate at zero magnetic field, but acquire different Zeeman energy shifts $E_{Z}$ in finite magnetic fields because their spin $z$-components (quantum number $m_S$) differ: $m_S\!=\!+1$ for $\ket{\:T_{+\!}}\!=\!\ket{\ua\ua}$, $m_S\!=\!0$ for $\ket{\:T_{0}}\!=\!(\ket{\ua\da}\!+\!\ket{\da\ua})/\sqrt{2}$ and $m_S\!=\!-1$ for $\ket{\:T_{-\!}}\!=\!\ket{\da\da}$.

The energies of the states can be expressed in terms of the single-particle energies of the two electrons plus a charging energy $E_{C}$ which accounts for the Coulomb interactions:
\begin{eqnarray*}
	&&\!\!\!\!\!\!\!\!E_S\ =\! E_{\ua}+ E_{\da} + E_{C} = 2 E_{\ua}+ \Delta E_{Z} + E_{C}\\
	&&\!\!\!\!\!\!\!\!E_{T_+}\!=\! 2 E_{\ua} + E_{ST} + \!E_{C}\\
	&&\!\!\!\!\!\!\!\!E_{T_0} =\! E_{\ua}\! + \! E_{\da}\! + \! E_{ST}\!+\!E_{C} = 2 E_{\ua}\!+\! E_{ST}\! +\! \Delta E_{Z}\! + \! E_{C}\\
	&&\!\!\!\!\!\!\!\!E_{T_-}\! =\! 2E_{\da}\!+\!E_{ST}\!+\!E_{C}=2 E_{\ua}\!+\!E_{ST}\!+\! 2 \Delta E_{Z} \!+\!E_{C},
\end{eqnarray*}
with $E_{ST}$ denoting the singlet-triplet energy difference in the absence of the Zeeman splitting $\Delta E_{Z}$.  $E_{ST}$ is considerably smaller than the single-particle level spacing $\Delta E_{orb}$, because the occupation of different orbitals and exchange interaction reduce the Coulomb energy for the triplet states \cite{LeoFewEl}.

\begin{figure}[t]
\includegraphics[width=3.4in]{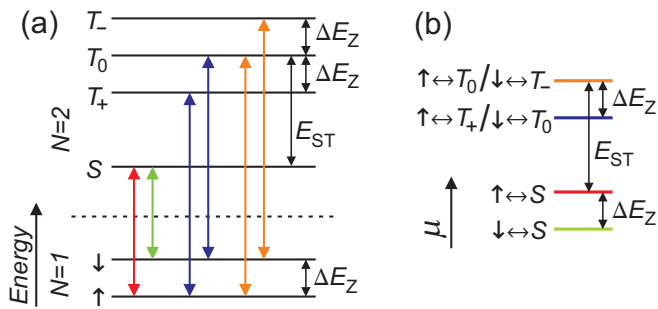}
\caption{(a) Energy diagram schematically showing the energy levels of the one- and two-electron states. The allowed transitions between these levels are indicated by arrows. (b) Electrochemical potential ladder corresponding to the transitions shown in (a), using the same color coding. Changing the gate voltage shifts the ladder as a whole. Note that the three triplet states appear at only two values of the electrochemical potential.}
\label{fig:energies}
\end{figure}

Fig. \ref{fig:energies}a shows the possible transitions between the one-electron spin-split orbital ground state and the two-electron states. We have omitted the transitions $\ua\leftrightarrow\!T_-$ and $\da\leftrightarrow\!T_+$, since these require a change in the spin $z$-component of more than $\frac{1}{2}$ and are thus spin-blocked \cite{Weinmann}. From the energy diagram we can deduce the electrochemical potential ladder, which is shown in Fig. \ref{fig:energies}b. Note that $\mu_{\ua \leftrightarrow T_+} = \mu_{\da \leftrightarrow T_0}$ and $\mu_{\ua \leftrightarrow \!T_0} = \mu_{\da \leftrightarrow \!T_-}$. Consequently, the \textit{three} triplet states change the first-order transport through the dot at only \textit{two} values of $V_{SD}$. The reason is that the first-order transport probes the energy difference between states with \textit{successive} electron number. In contrast, the onset of second-order (cotunneling) currents is governed by the energy difference between states with the \textit{same} number of electrons. Therefore, the triplet states change the second-order (cotunneling) currents at three values of $V_{SD}$. In our measurements, these cotunneling currents were too small to detect (see below).

\begin{figure}[t]
\includegraphics[width=3.1in]{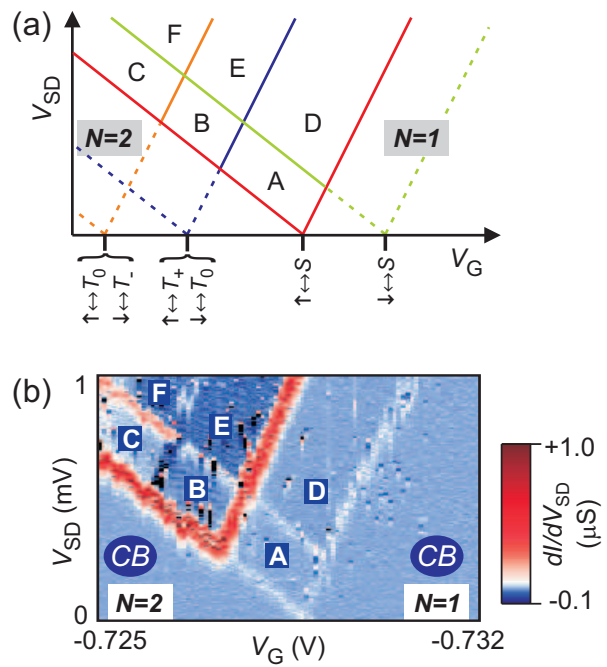}
\caption{(a) Energetically allowed 1$\leftrightarrow$2 electron transitions as a function of $V_{SD}$ and $V_{G}$. The lines corresponding to $\ua\leftrightarrow\!S$ outline the region of transport; outside this region, where lines are dashed, the dot is in Coulomb blockade. (b) $dI/dV_{SD}$ as a function of  $V_G$ and $V_{SD}$ around the 1$\leftrightarrow$2 electron transition at $B_{/\!/}\!=\!$ 12 T. In the region labeled $A$ only spin-down electrons pass through the dot.}
\label{fig:N=2}
\end{figure}

In Fig. \ref{fig:N=2}a we map out the positions of the electrochemical potentials as a function of $V_G$ and $V_{SD}$. For each transition, the two lines originating at $V_{SD}\!=\!$ 0 span a V-shaped region where the corresponding electrochemical potential is in the bias window \cite{DiamondSkewness}. In the region labeled $A$, only transitions between the one-electron ground state, \ket{\ua}, and the two-electron ground state, \ket{\:S}, are possible, since only $\mu_{\ua \leftrightarrow S}$ is positioned inside the bias window. Since this transition corresponds to transport of spin-down electrons \textit{only}, the dot again acts as a spin filter, but with a polarization opposite to the 0$\leftrightarrow$1 electron case. The polarization of the current is lost when $\mu_{\da\leftrightarrow S}$ or $\mu_{\ua \leftrightarrow T_+}$ enters the bias window (regions $D$ an $B$ respectively). In the regions $C,E$ and $F$ several more transitions are possible which leads to a more complex, but still understandable behavior of the current. Outside the V-shaped region spanned by the ground state transition $\mu_{\ua\leftrightarrow S}$, Coulomb blockade prohibits first order electron transport.

%N=2 DATA%%%%%%%%%%%%%%%%%%%%%%%%%%%%%%%%%%%%%%%%%%%%%%%%%

The experimental data, shown in Fig. \ref{fig:N=2}b, is in excellent agreement with the predictions of Fig. \ref{fig:N=2}a. Three important observations can be made. First, we clearly observe the Zeeman splitting of the triplet states. Second, the transitions between the one-electron states and the two-electron singlet ground state are spin-resolved. Thus, we can easily identify the region where the current is carried by spin-down electrons only, i.e. region $A$. The size of this region is determined by the Zeeman energy ($\Delta E_Z\!\approx\!$ 240 $\!\mu$eV) and the singlet-triplet energy difference ($E_{ST}\!\approx\!$ 520 $\!\mu$eV). The third observation is that the Zeeman energy, and therefore the $g$-factor, is the same for the one-electron states as for the two-electron states, within measurement accuracy ($\approx\!5\%$). We note that the large variation in differential conductance observed in Fig. \ref{fig:N=2}b, can be completely explained by a sequential tunneling model with spin- and orbital-dependent tunnel rates \cite{Moriond2004}.

To achieve spin-down filtering at the 1$\leftrightarrow$2 electron transition, it is crucial that the two-electron ground state is a spin-singlet. Indeed, in case of a triplet ground state, the dot would transmit only spin-up electrons. We made sure that the ground state of the two-electron dot at $B_{/\!/}\!=\!$ 12 T is still a spin-singlet state by carefully monitoring the energies of the two-electron states from zero field, where the ground state is always a spin-singlet \cite{Ashcroft}. Additionally, the line of high $dI/dV_{SD}$ separating region $A$ from $D$ would not be present in case the ground state would be a spin triplet. (This can be seen by redrawing the diagram in Fig. \ref{fig:N=2}a for the case of a triplet ground state).

%COTUNNELING %%%%%%%%%%%%%%%%%%%%%%%%%%%%%%%%%%%%%%%%%%%%%%%%%
The data presented in Figs. \ref{fig:N=1data}d and \ref{fig:N=2}b shows that our device can be operated as a bipolar spin filter, as proposed by Recher \textit{et al}. Switching between the 0$\leftrightarrow$1 electron transition, where the polarization is spin-up, and the 1$\leftrightarrow$2 electron transition, where the polarization is spin-down, only requires adjusting the gate voltages, which can already be done on a subnanosecond timescale \cite{ZeemanRel}.

In a sequential tunneling picture, the polarization of the first-order tunnel current is, due to energy conservation, $\approx$100\% whenever $\Delta E_Z \gg kT$ (which is easily fulfilled here). We now investigate the influence of tunneling via virtual higher-energy states (second-order tunneling or cotunneling) \cite{Nazarov} and of spin-orbit coupling on the filter efficiency. 

We first note that the cotunneling current $I_{cot}\!\propto\!\Gamma^2$, whereas the first-order tunneling current $\propto\!\Gamma$ and therefore cotunneling can always be suppressed by making the tunnel rates small. We obtain an upper bound on the cotunneling current by measuring the current in the Coulomb blockade region close to the spin filter region. Here, the parameters for cotunneling are the same as those in the spin filter region, but first-order tunneling is forbidden, allowing a direct measurement of $I_{cot}$. We find that for both the 0$\leftrightarrow$1 and the 1$\leftrightarrow$2 electron transition, $I_{cot}$ is smaller than the noise floor of our measurement ($10^{-14}$ A).

Using values for the tunnel rates obtained from first-order tunneling, we can also calculate $I_{cot}$ in the spin filter regions. For $V_{SD}<\Delta E_Z$, only elastic cotunneling is possible \cite{Nazarov}. At the 0$\leftrightarrow$1 electron transition, we find that $I_{cot}$ is $\sim\!10^{-19}$A in the middle of region I, whereas the sequential spin-up current is $\sim\!10^{-13}$A. At the 1$\leftrightarrow$2 electron transition, $I_{cot}\!\sim\!10^{-15}$A in the middle of region $A$. With $\mu_{\ua\leftrightarrow T_+}$ far above $\mu_S$, this reduces to $10^{-16}$A, which is more than three orders of magnitude smaller than the sequential spin-down current ($\approx\!3\!\cdot\!10^{-13}$A). Thus, both the measurements and the calculations show that second-order tunneling processes are negligible.

Due to spin-orbit coupling the eigenstates in the dot are not the pure spin states \ket{\ua} and \ket{\da}, but each contains a small admixture of the opposite spin, which limits the efficiency of the spin filter in the ($\ua$,$\da$) basis. An upper bound on the spin-orbit coupling can be derived from the spin-orbit mediated spin relaxation. The very low spin relaxation rates measured in our device, 2 MHz at 10 T and 9 MHz at 14 T \cite{ZeemanRel,NatureReadout}, indicate that the reduction in efficiency is less than $10^{-4}$ \cite{Golovach}. We further note that the tunnel barriers are purely electrostatically defined and should therefore not induce extra spin relaxation.

Future experiments will focus on measurement of the spin-polarization of the current flowing out of the dot by an external analyzer. This can be done for instance by focussing the current onto a quantum point contact \cite{JoshScience}, although this technique has only allowed polarizations up to ~70\% to be measured. Alternatively, we plan to investigate the filter properties by placing two dots in series \cite{Ono}, such that the polarization configuration can be switched controllably between parallel and antiparallel.

%ACKNOWLEDGEMENTS
We thank D.P. DiVincenzo, C.M. Marcus, T. Fujisawa, S. Tarucha, T. Hayashi, T. Saku, Y. Hirayama, A. Sachrajda, J.A. Folk,  D. Loss, V.N. Golovach and R.N. Schouten for discussions and help. This work was supported by the DARPA-QUIST program, the ONR, FOM and the EU-RTN network on spintronics.


\begin{references}
\bibitem{DiamondSkewness}
We apply the source-drain bias asymmetrically; the drain is connected to ground in the \textit{I-V} converter. The skewness of the V-shaped transport region is determined by the ratio of the capacitances of the dot to the source, $C_S$, and to the drain, $C_D$. For $C_S$=$C_D$, the V-shape is symmetric.
\end{references}
\end{document}